# Realization of Woodpile Structure Using Optical Interference Holography

*Yee Kwong Pang, Jeffrey Chi Wai Lee, Cheuk Ting Ho and Wing Yim Tam\**

Department of Physics and Institute of Nano Science and Technology
Hong Kong University of Science and Technology
Clear Water Bay, Kowloon, Hong Kong, China

Photonic crystals are dielectric materials that exhibit bandgaps in which EM waves propagation in the bandgaps are forbidden.[1] To achieve photonic crystals with complete bandgaps has been challenging to both theorists and experimentalists.[1,2] Among the various structures that supports complete photonic band gaps, the diamond and the related woodpile structures stand out with wide and robust bandgaps even with a moderate dielectric contrast.[2] Various techniques, including the self-assembly or nano-manipulation of colloidal microspheres,[3,4] the layer by layer micro-fabrications,[5] and recently, the holographic lithography[6-8] and the multi-photon direct laser writing,[9] have been used to fabricate photonic crystals. However, not all techniques are suitable for the woodpile/diamond structure. For example, the self-assembly method was limited to face-centered-cubic (FCC) or close-packed structures.[3] Recently, micro-manipulation has been used to fabricate the diamond structure.[4] However, the sample size is limited to a few unit cells and the process is very tedious. The layer by layer and the multi-photon direct laser writing techniques had been used to fabricate the woodpile structure with bandgaps in the infrared range.[5,9] However, it was limited to a few layers in the first case, and was tedious for both due to the demanding precisions and procedures.[5,9] The holographic lithography, a method combining the techniques of multiple beams interference and photolithography to record the interference pattern in photo-resist, provides some unique advantages. For example, it requires only simple experimental setups, and more importantly, various structures (e.g. quasi-periodic[7] and chiral structures[8]) are feasible by varying the beam orientations and polarizations. This method has thus attracted much interest since the realization of the FCC structure using the interference of four non-planar coherent beams.[6] Furthermore a double-exposure holographic lithography had also been used to fabricate the woodpile structure in the infrared range.[10] Recently, several groups have suggested that the diamond structure could be fabricated using 4-beam configurations.[11-14] However, these configurations require either impractical beam arrangements or elliptical polarizations that are hard to implement experimentally.[11-14] A recent attempt to fabricate the diamond structure using a (3+1)-beam configuration (3 linear polarized side beams and one circular polarized central beam), to simulate the double exposures for two FCC structures, was debatable.[16,17] One of us has proposed recently that the woodpile and diamond structures can be obtained using a 5-beam optical interference holography that is accessible experimentally all the beams are from the same half space as compared to other configurations in which the interfering beams are counter-propagating from both half spaces.[17] In this communication, we report the use of a (4+1)-beam interference configuration to fabricate the woodpile structure in photo-resist using one single exposure. The configuration is basically the "umbrella" arrangement with 4 linear polarized side beams arranged symmetrically around a circular polarized central beam.[17] The fabricated woodpile structures, in submicron scales, are in good agreement with model simulations. Furthermore, they also exhibit directional bandgaps in the visible range.

The wave vectors of the beams for the woodpile structure as shown in Fig. 1(a) can be represented by[17]

$$\begin{bmatrix} \vec{k}_0 = k(0,0,1) \\ \vec{k}_1 = k(-\sin\varphi, 0, \cos\varphi) \\ \vec{k}_2 = k(0, -\sin\varphi, \cos\varphi) \\ \vec{k}_3 = k(\sin\varphi, 0, \cos\varphi) \\ \vec{k}_4 = k(0, \sin\varphi, \cos\varphi) \end{bmatrix}, \quad (1)$$

where $k = 2\pi/\lambda$ for $\lambda=488nm$, the wavelength of the light source. Here $\varphi$ is the angle between the side beams $\vec{k}_i$ and the central beam $\vec{k}_0$. The central beam is circular polarized with electric field given by $\vec{E}_0 = \frac{E_0}{\sqrt{2}}(1, i, 0)$ while the side beams are linear polarized with electric fields normal to the plane of incidence. Given Eq. (1), the intensity distribution of the (4+1)-beam interference can be expressed as[17]

$$I(\vec{r}) = \sum_{l,m} \vec{E}_l \cdot \vec{E}_m^* e^{-i\vec{q}_{lm} \cdot \vec{r} - i(\delta_l - \delta_m)}, \qquad (2)$$

where $\vec{q}_{lm} = \vec{k}_l - \vec{k}_m$ for $l,m$ = 0-4 and $\delta$'s are the phases of the beams. For simplicity, we choose $|\vec{E}_i|$ =1 for i =0-4. Figure 1(b) shows a woodpile structure obtained by the superposition of orthogonal x- and y- directional rods stacked and interlaced with half rod-space shift in each plane. The x- and y- rods are obtained by the interference of beams $(\vec{k}_0, \vec{k}_2, \vec{k}_4)$ and $(\vec{k}_0, \vec{k}_1, \vec{k}_3)$, respectively. The rod spacings ($a$ for x- and $b$ for y- rods as shown in Fig. 1(b)) and the shape of the rods depend on the angle $\varphi$. For diamond symmetry, the lattice ratio $a/b$ is equal to $1/\sqrt{2}$, corresponding to $\varphi = 70.53°$. Figures 1(c) and (d) show woodpile structures as intensity contour surfaces obtained by Eqs. 1 and 2 for all beams with the same phases ($\delta$'s =0) and one beam 180° out of phase (e.g. $\delta_2$= 180° and others are zero). To compare with the experimental results, the structures are simulated using the experimental incidence angle $\varphi = 41.8°$ taking the ~40% shrinkage in the z-direction and ~10% expansion in the xy-directions into account to achieve the diamond symmetry. One obvious difference between Figs. 1(c) and (d) is that the x- and y- rods in Fig. 1(c) are correctly stacked and interlaced while they occupy the same z-position in each plane for the incorrect phase case as shown in Fig. 1(d). The top views (upper-right insets in Figs. 1(c) and (d)) further demonstrate the stacking of the rods in both cases. The lower-right inset of Fig. 1(c) obtained at a higher intensity cut-off, shows clearly a diamond structure for the correct phase configuration while the corresponding inset in Fig. 1(d) shows a FCC structure interlaced with z-directional rods in between. For higher intensity cut-offs, the z-rods will disappear giving only a FCC structure. Fortunately, good woodpile/diamond structure can still be obtained for small phase difference as confirmed by 200 realizations of random-phase simulations within which more than 50% still show visually discernible woodpile structure.

The realization of the woodpile structure was carried out by using a 5-beam holographic lithography technique on a photo-resist. Figure 2(a) shows an SEM image of a sample fabricated as described in the experimental section. The sample consists of more than 20 layers of rods resemble strikingly to that obtained from the model. We found that the samples had shrank/collapsed substantially, ~40%, in the z-direction while there was about a 10% change in the xy-directions. The shrinkage was anticipated from past experience and was the main reason we selected a smaller incident angle instead of the expected angle $\varphi = 70.53^o$ for the diamond structure.[7,8] Figures 2(b) and (c) show woodpile structures with lattice ratio (obtained from the front view insets) $a/b$ = 0.72 and 0.88. They look very similar to that obtained by the multi-photon direct laser writing technique.[9] The main advantage of our method is that it takes much shorter time than that used in the direct laser writing technique. In addition, large samples of a few mm can be fabricated even though uniform regions with the correct phases are small due to inhomogeneities in the optics. Note that Fig. 2(b) is very close to the diamond symmetry within experimental uncertainty while Fig. 2(c) differs slightly by a larger shrinkage in both the x- and y- directions than those in Fig. 2(b). Figure 2(d) shows an example with lattice ratio $a/b$ = 0.82 for the most unfavourable phases for the diamond lattice. It is clear that the x- and y- rods are all in the same plane at each layer, resembling closely to Fig. 1(d). The front view inset of Fig. 2(d) shows clearly the different stacking as compared to the samples with the correct phases as shown in the front view insets of Figs. 2(b) and (c). The agreement between the model the experiment can also been seen from the top view (lower-left) insets of Figs. 2(b)-(d) and the upper-right insets of Fig. 1(c)-(d) for both the correct (Figs. 1(c) and 2(b)-(c)) and incorrect (Figs. 1(d) and 2(d)) phases.

Despite the differences in the rod stacking, all samples possess visible range directional bandgaps in the normal reflectance and transmittance measured as described in the experimental section and shown in Figs. 3(a)-(c), corresponding to Figs. 2(b)-(c) respectively. There is a bandgap around 700 nm, shifting slightly to longer wavelength, for the correct rod stacking samples in Figs. 2(b)-(c) respectively. Despite the incorrect stacking in Fig. 2(d), the bandgap is also ~ 700nm suggesting that either our samples are not good enough or the normal incident is not sensitive enough to distinguish the difference. Unfortunately, our samples do not have uniform regions large enough for obtaining reliable angular dependent measurements and to warrant a quantitative comparison with calculations.

To conclude, we have fabricated the woodpile structure on photo-resist using a (4+1) beam optical interference holography. The samples resemble the simulations very well. By exploiting the shrinkage of the photo-resist, samples with the diamond lattice spacing is obtained. The woodpile structures display visible range directional bandgaps. Our samples, although small in size and have low dielectric contrast, could be used as template for fabricating woodpile structures with higher dielectric contrast for complete bandgaps.

*Experimental*

The five beams, 7.5 mm diameter, in Fig. 1(a) were obtained by passing an expanded beam from an argon ion laser through a template with one central hole and four side holes distributed evenly around the central hole. The beams, power 4.5 mW each and polarizations adjusted by wave plates mounted at the holes of the template entered a four-sided truncated pyramid from the base as shown in Fig. 1(a). The central beam, converted to circular polarized by a quarter wave plate, went straight up the pyramid while the side beams reflected internally at the slanted surfaces and intersect at the truncated surface making an angle $\varphi = 41.8^{\circ}$ with the central beam as shown in Fig. 1(a). Using this setup, the beams were more uniform and, most importantly, the phases of the beams were fixed because they were obtained from the same expanded beam. We used the photo-resist "SU8" (from Shell) as the raw polymer resin and followed the procedures reported earlier.[7,8] The resin was spin-coated on glass substrates with almost the same refractive index as the SU8 to form ~20 μm thick samples. The samples were heated to 90$^{o}$C to remove any solvent left before exposure. The photo-resist coated sample was placed on the truncated surface of the pyramid with index-matching to reduce multiple reflections. The exposure time was 15 s. After the exposure, a post-thermal treatment at 90$^{o}$C for about 30 mins was needed to complete the polymerization. Polymerization occurred only at regions where the dosage exceeded a critical value, while under-exposed regions were washed away first by bathing the sample with propylene-glycol-methyl-ether-acetate (PGMEA) for 8 hours, then rinsed with PGMEA-acetone solution, and finally with ethanol, creating a copy of the woodpile pattern. We obtained the normal reflection and transmission spectra by using an optical microscope coupled to a spectrometer (Oriel Cornerstone 260) through an optical fiber as reported recently.[7] The microscope, with a pin hole installed in the optical path, could be focused down to a size of 15 μm using a 100X objective. Reflectance and transmittance were normalized against backgrounds of silver mirror reflection and empty air transmission, respectively. SEM images were obtained by using a JEOL 6300F scanning electron microscope.

Keywords: photonic crystals, woodpile and diamond structures, holography

* Corresponding Author: phtam@ust.hk; Phone: 852-2358-7490; Fax: 852-2358-1652.

**Acknowledgment**

We thank C. T. Chan for helpful discussions. Support from Hong Kong RGC grants CA02/03.SC01, HKUST603303, and HKUST603405 is gratefully acknowledged.

**Figure captions**

1) (a) 5-beam configuration for the woodpile structure. (b) Superposition of x- rods and y- rods obtained by the interference of $(\vec{k}_0,\vec{k}_2,\vec{k}_4)$ and $(\vec{k}_0,\vec{k}_1,\vec{k}_3)$, respectively. (c) Woodpile structure shown as intensity contour surfaces with a 50% cut-off by the interference of $(\vec{k}_0,\vec{k}_1,\vec{k}_2,\vec{k}_3,\vec{k}_4)$ beams with equal phases and using $\varphi = 41.8°$. The structure is compressed by 40% and expanded by 10% along the z and in the xy-directions, respectively, to simulate the deformations observed in the experiment. (Same result is obtained using $\varphi = 70.53°$ but without the deformations as reported in Ref. [17].) The insets, upper-right (50% cut-off) and lower-right (95% cut-off), are views of the top and the unit cell of the diamond structure, respectively. (d) Contour surfaces with a 40% intensity cut-off for the 5-beam interference similar to (c) but with $\vec{k}_2$ $180°$ out of phase w.r.t. the other beams. The insets, upper-right (40% cut-off) and lower-right (60% cut-off), are views of the top and the unit cell, respectively.

2) (a) 3D SEM images of woodpile structure. The upper-left inset shows the expanded view of the woodpile structure. (b) – (d) SEM images for the woodpile structures with $a/b$ =0.72, 0.88, and 0.82, respectively. Note that (b) and (c) show the favorable results with the x- and y- rods properly interlaced while (d) shows the unfavorable result with the x- and y- rods in the same plane for each layer. The upper-right insets (size 1.5x1.0μm$^2$) are the expanded front views while the lower-left insets (size 1.8x1.8μm$^2$) are the top views of the structures. The scale bars (white) are all 1.0 μm.

3) (a)-(b) Normal reflectance (in blue) and transmittance (in red) for the samples in Fig. 3(b)-(d), respectively.

Fig. 1

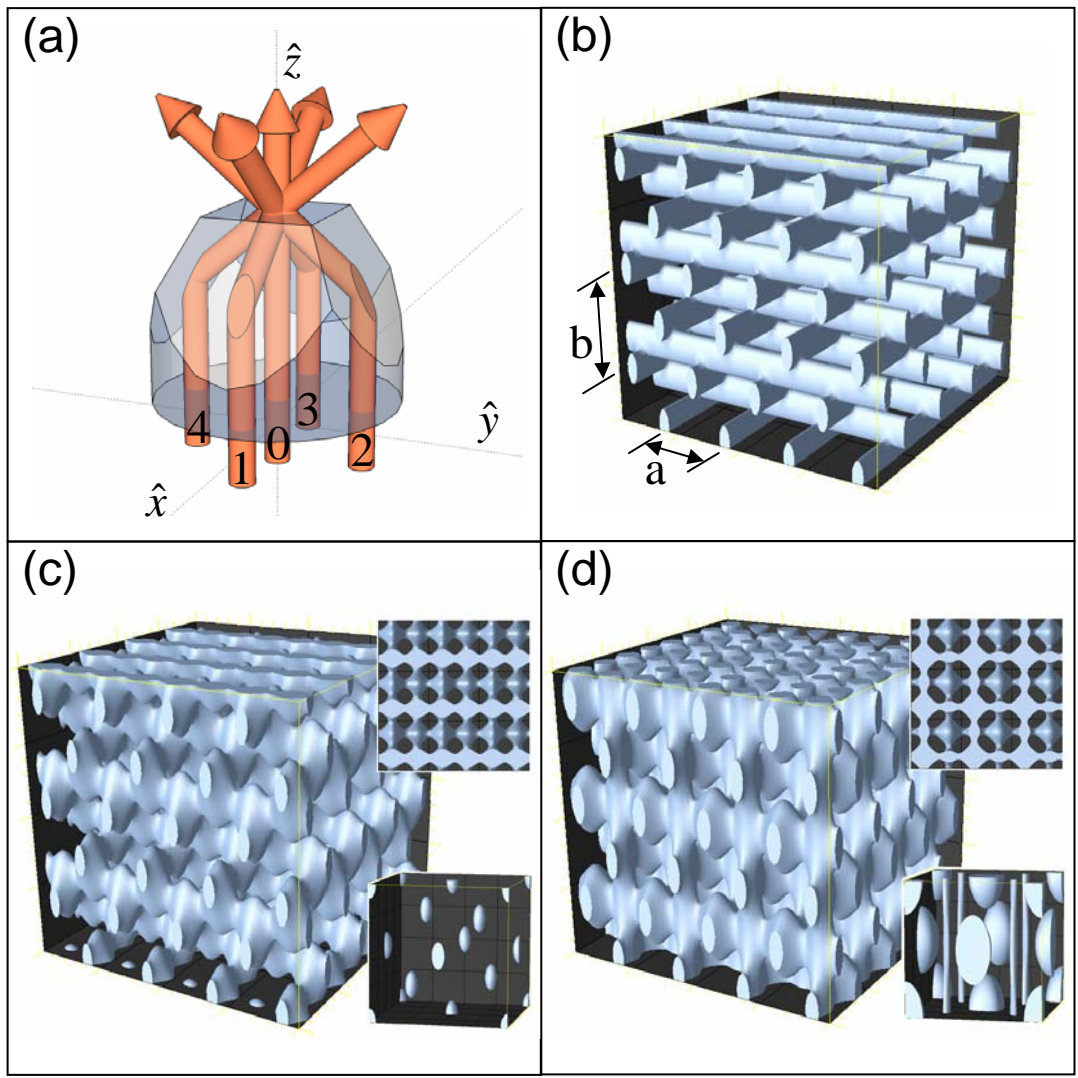

Fig. 2

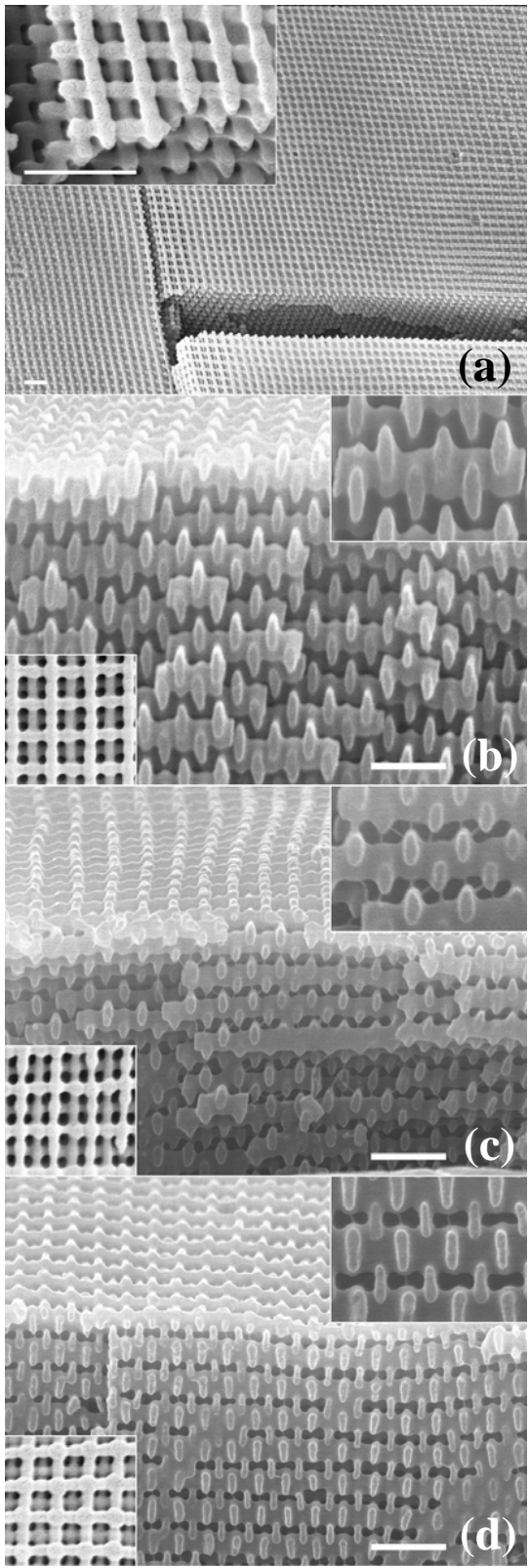

Fig. 3

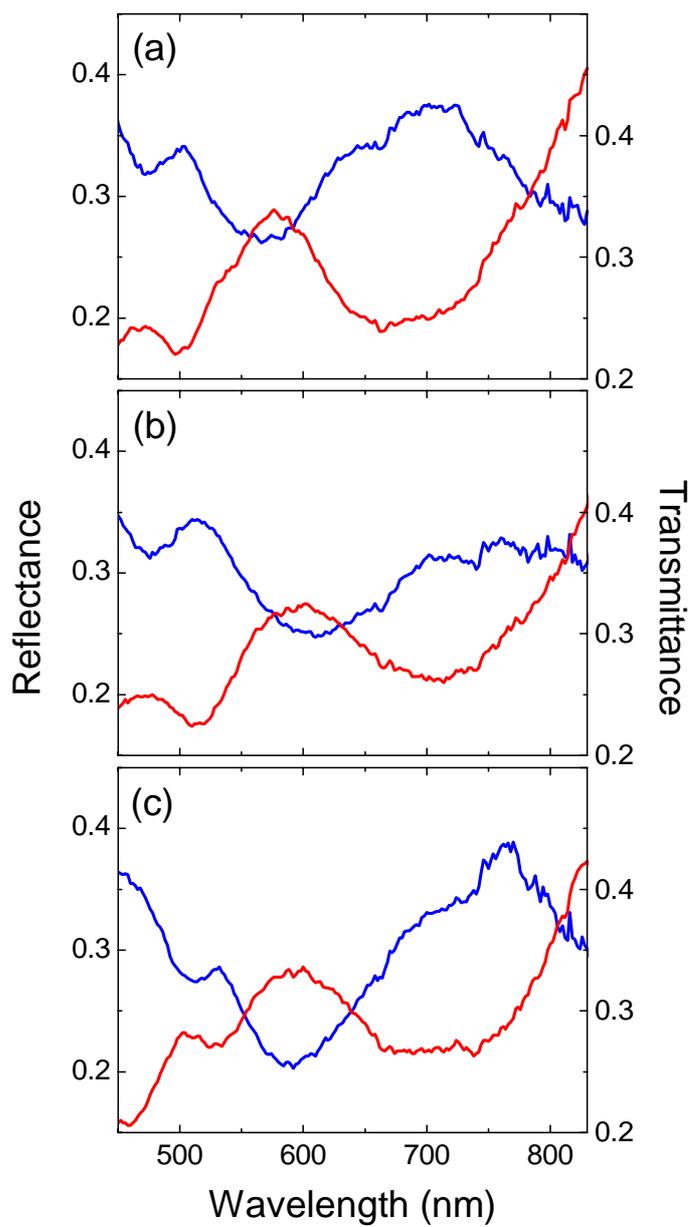